\documentclass[twocolumn,aps]{revtex4b4}

\usepackage{graphicx}

\begin{document}

\title{Resistance Spikes at Transitions between Quantum Hall Ferromagnets}
\author{E. P. De Poortere, E. Tutuc, S. J. Papadakis, and M. Shayegan}
\address{Department of Electrical Engineering, Princeton University, Princeton, New Jersey 08544}
\date{\today}

\begin{abstract}
We report a manifestation of first-order magnetic transitions in two-dimensional electron systems. This phenomenon
occurs in aluminum arsenide quantum wells with sufficiently low carrier densities and appears as a set of
hysteretic spikes in the resistance of a sample placed in crossed parallel and perpendicular magnetic fields, each
spike occurring at the transition between states with different partial magnetizations. Our experiments thus
indicate that the presence of magnetic domains at the transition starkly increases dissipation, an effect also
suspected in other ferromagnetic materials. Analysis of the positions of the transition spikes allows us to deduce
the change in exchange-correlation energy across the magnetic transition, which in turn will help improve our
understanding of metallic ferromagnetism.
\end{abstract}

\maketitle

Ferromagnetism in metallic systems, also known as itinerant electron ferromagnetism, has thus far revealed few of
its secrets to scientists. In one of its well-known occurrences, in transition elements Fe, Co, and Ni, metallic
ferromagnetism is believed to stem from a partially filled 3d band of electrons with unbalanced spin populations,
although the properties of their magnetic moments at nonzero temperature are still unclear {\it (1)} despite some
recent theoretical progress {\it (2)}. In dilute electron gases, the appearance of spontaneous magnetization at
sufficiently low densities, as evidenced by recent experiments in doped hexaborides {\it (3)}, is also subject to
debate, with the critical density at the ferromagnetic to paramagnetic transition still uncertain {\it (4)}. One
difficulty in modeling these materials lies in their complexity, because itinerant charged carriers in these
systems are subject not only to electron-electron interactions, but also to atomic potentials resulting in an
intricate density of states or to local moments of other atoms in the material.

Two-dimensional (2D) electron systems in modulation-doped semiconductor heterostructures, on the other hand,
provide an ideal system for the study of itinerant electron ferromagnetism, as interactions with their host
material are almost entirely contained in the effective mass ($m^*$) and effective $g$-factor ($g^*$) of
electrons. Given these two renormalizations, carriers in these structures behave as a nearly free electron gas. In
a perpendicular magnetic field ($B_\bot$), 2D electrons condense into a ladder of energy levels, called Landau
levels, which are separated by the cyclotron energy $\hbar \omega_c = \hbar e B_\bot/m^*$. The total magnetic
field ($B_{tot}$) also couples to the electron spin, and leads to an additional (Zeeman) energy $\pm {1 \over
2}|g^*|\mu _BB_{tot}$ (where $\mu_B$ is the Bohr magneton, $e\hbar/2m_e$, and $m_e$ is the bare electron mass),
which splits each Landau level into two separate levels. The number of occupied spin-split Landau levels at a
given field is called the filling factor, $\nu$. This discrete level structure gives rise to the quantum Hall (QH)
effect, the vanishing of longitudinal resistance ($R_{xx}$) and the quantization of the Hall resistance, that
occurs when an integral number of Landau levels are occupied. In the setup we use for our experiments (inset of
Fig.\ 1B), where the sample is tilted at angle $\theta$ with respect to the magnetic field, both Zeeman and
cyclotron energies can be tuned independently.

\begin{figure*}
\includegraphics[width=15cm]{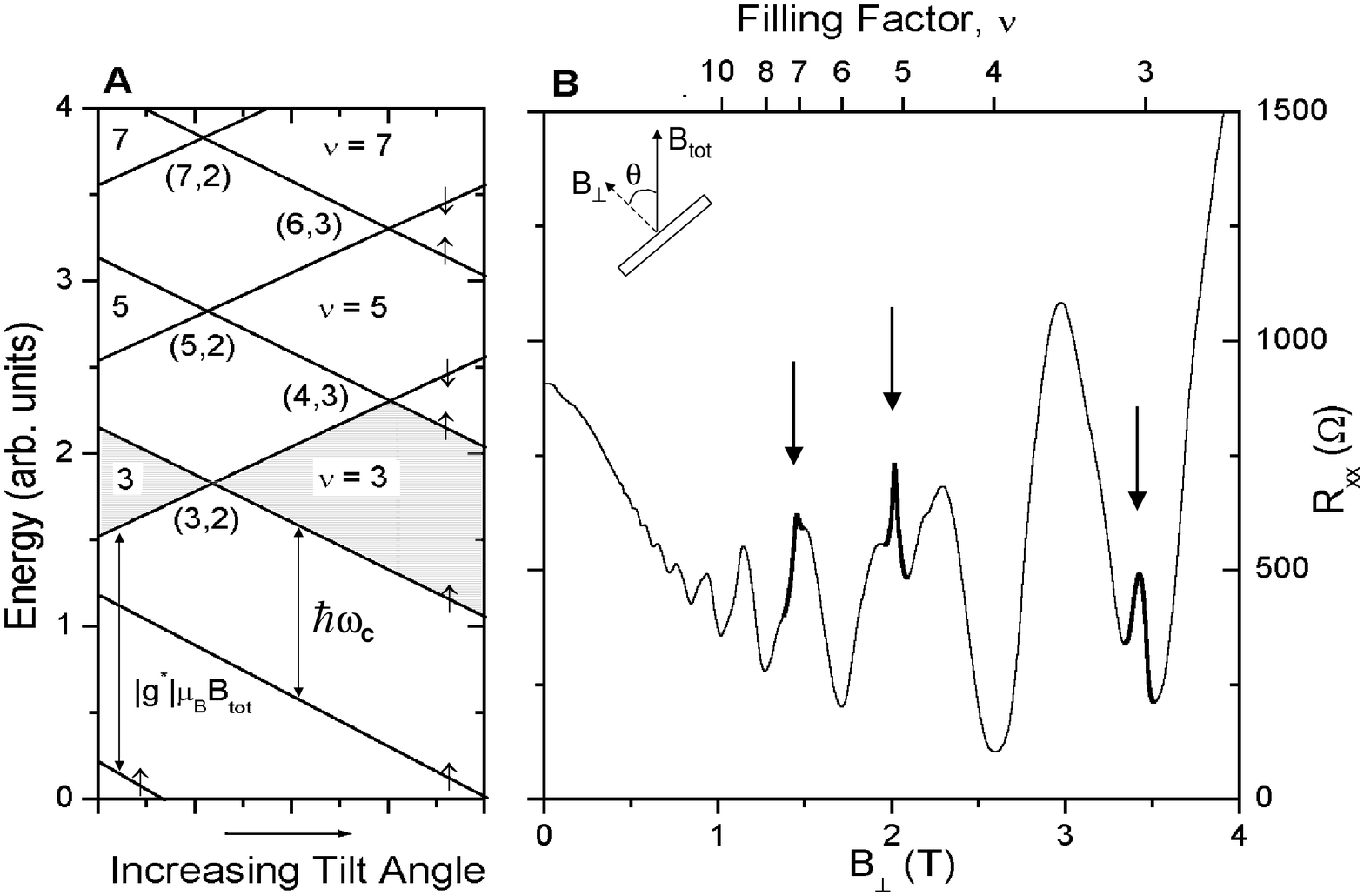}
\caption{{\bf(A)} Diagram of the Landau energy levels in a 2D electron gas, showing the spin-up and spin-down
levels intersecting as parallel field (or angle $\theta$) is increased. Indices ($\nu$,$l$) label each crossing,
where $\nu$ maps to the filling factor of the nearest magnetoresistance minimum and $l$ is the number of fully
polarized filled levels at the ($\nu$,$l$) coincidence. {\bf (B)} Longitudinal magnetoresistance ($R_{xx}$) of the
2D electron gas in our AlAs quantum well at $\theta = 38^o$ and at $T \approx 300$ mK, showing resistance spikes
near filling factors 3, 5 and 7.}
\end{figure*}

Figure 1A depicts the evolution of electronic levels in a constant $B_\bot$, as $B_{tot}$ (or $\theta$) is
increased: The Zeeman splitting increases with $B_{tot}$, so that in an independent-electron picture, energies
cross or come into ``coincidence'' for particular values of $\theta$. On either side of these coincidences, the
magnetization of the electron system thus takes on distinct values, corresponding to QH ferromagnets (or ``Ising''
states) of various strengths. Figure 1B highlights our main finding, namely that the transitions between these
ferromagnets provoke sharp peaks in the magnetoresistance. These new peaks, marked by arrows, are distinct from
the usual maxima between QH minima and are visible here near $\nu =$ 3, 5, and 7. Moreover, as we show later in
this report, the magnitude of the spikes at low temperatures depends strongly on magnetic field sweep direction, a
clear indication that the peaks occur at the magnetic phase transition. In addition, we are able to record these
peaks for a broad range of filling factors and carrier densities. This allows us to establish a simple model for
the field at which the transitions take place and, with this model, to obtain the exchange-correlation energy
gained by the system as it makes the transition.

Our results can be compared with tilt experiments performed in various 2D systems {\it (5-8)}. These previous
measurements either show the expected disappearance of the QH resistance minimum at coincidence, corresponding to
the vanishing of the gap between Landau levels, or, as in {\it (6, 7)}, reveal the persistence of the $R_{xx}$
minimum at coincidence, indicating a more subtle anticrossing of the Landau levels at the Ising transition. Our
experiment, on the other hand, displays detailed structure within the QH minimum: The latter remains strong over
most of its range, except in a small field interval where the resistance rises sharply. Fractional QH states can
also undergo a spin-related phase transition {\it (9-12)}, which appears to share common traits with the integer
QH transition. A theoretical investigation of magnetism and magnetic transitions in the integer QH regime has been
addressed in {\it (13)}, motivated in part by experiments on pseudospin Ising transitions in wide GaAs quantum
wells (the pseudospin of an electron refers to its layer or subband index) {\it (8, 14)}.

The 2D electron gas we study resides in the conduction band of a 150\AA -wide AlAs quantum well, bordered by
Al$_{0.39}$Ga$_{0.61}$As barriers and modulation-doped with a front layer of Si. Our structures were grown by
molecular-beam epitaxy over GaAs substrates oriented along various crystal lattice planes, and the data we present
are from a sample grown on a (411)B-oriented substrate. We performed magnetotransport measurements in pumped
$^3$He and dilution refrigerators with magnetic fields up to 18T. Through a combination of sample illumination and
back-gate biasing, we were able to vary the carrier density ($n$) from $1.6 \times 10^{11}$ to $5.2 \times
10^{11}$ cm$^{-2}$.

\begin{figure*}
\includegraphics[width=17cm]{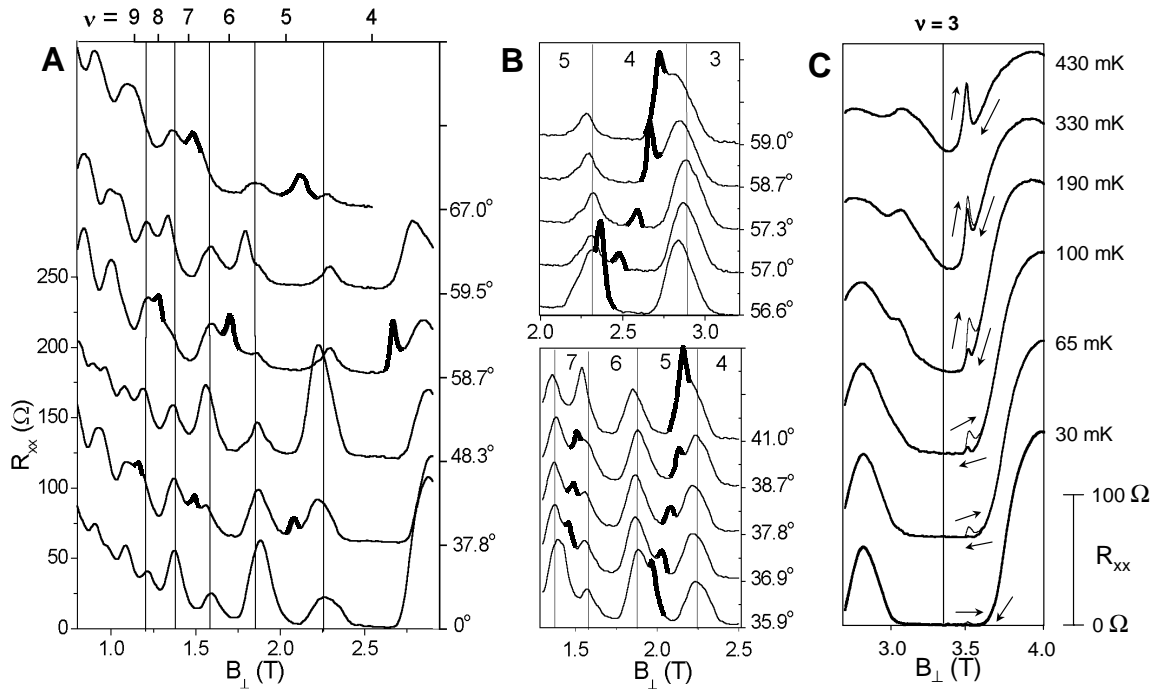}
\caption{{\bf (A)} Magnetoresistance of the 2D electron gas at $n = 2.5 \times 10^{11}$ cm$^{-2}$ for various tilt
angles at $T = 30$ mK. New peaks are seen near $\nu =$ 5, 7, and 9 for $\theta = 37.8^o$ and $\theta = 67.0^o$,
and near $\nu =$ 4, 6, and 8 for $\theta = 58.7^o$. $R_{xx}$ traces are shifted vertically for clarity. Base lines
($R_{xx} = 0$) are marked by the right tick marks, which indicate the corresponding tilt angle. {\bf (B)} Movement
of the peaks from higher to lower filling as tilt angle increases. {\bf (C)} Hysteresis and temperature dependence
of the magnetoresistance peak near $\nu = 3$, at a tilt angle = 37.8$^o$. No hysteresis is seen at 430 mK, whereas
at lower temperatures the peak becomes increasingly hysteretic. At 30mK, the peak has all but vanished in the
downward sweep, whereas it is still observable (in an enlarged trace) in the upward sweep.}
\end{figure*}

Our choice of AlAs as a host material for 2D electrons stems from the magnitude of their effective mass {\it (15)}
and of their $g$-factor, both of which are substantially larger than in GaAs. The larger mass in AlAs makes
carriers more dilute and therefore more sensitive to many-body effects; the larger $g$-factor, equal to 1.9 {\it
(16)}, implies that the Zeeman and cyclotron energies become comparable for a wide range of experimental
conditions. Moreover, our samples have electron mobilities as high as 20 m$^2$/Vs for $n = 5 \times 10^{11}$
cm$^{-2}$, and exhibit fractional QH states up to third-order fractions, indicating that many-body effects
dominate the energetics of our 2D electron system at low temperatures.

We studied the behavior of the anomalous peaks while changing various parameters: tilt angle, temperature, field
sweep direction, and sample density. Fig.\ 2, A and B, shows the evolution of magnetoresistance as we change the
tilt angle of our sample (here $n = 2.5 \times 10^{11}$ cm$^{-2}$). At three angles, 37.8$^o$, 58.7$^o$ and
67.0$^o$, additional peaks appear within the QH minima.  At $\theta = 37.8^o$ and $\theta = 67.0^o$, peaks are
seen near odd fillings $\nu =$ 5, 7, and 9, and for $\theta = 58.7^o$ they occur near even fillings $\nu = 4, 6$,
and $8$. Moreover, as we increase $\theta$ around these special angles (Fig.\ 2B), we see the peaks travel from
low fields to high fields. This allows us to rule out an inhomogeneous density in the sample as a possible
explanation for the peaks, because inhomogeneities would not depend on the parallel field.

Evidence that the anomalous peaks correspond to a phase transition is displayed in Fig.\ 2C: Tuning the 2D
electrons to the state where a peak is seen near $\nu = 3$, we observe magnetic hysteresis at the precise location
of the anomalous peak. At high temperatures ($T = 430$ mK), hysteresis is absent, whereas for $T < 330$ mK, upward
and downward sweeps diverge increasingly as temperature is lowered. For $T < 65$ mK the downsweep peak cannot be
discerned, and likewise at $T = 30$ mK, the upsweep peak has nearly vanished.

\begin{figure*}
\includegraphics[width=17cm]{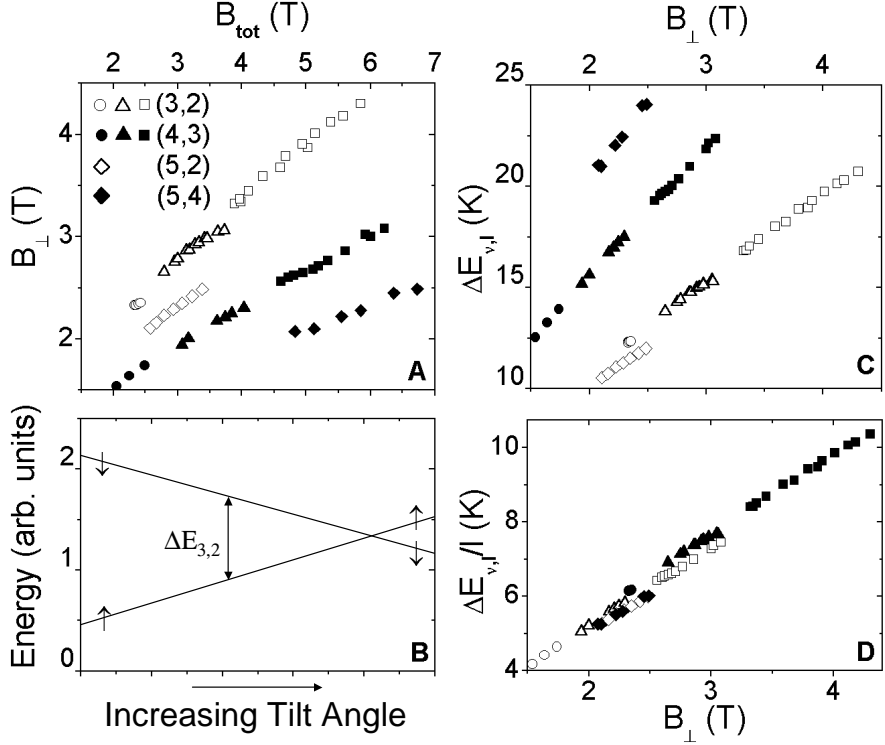}
\caption{{\bf (A)} Positions of the magnetoresistance peaks, plotted in ($B_\bot$, $B_{tot}$) space for various
densities (squares/diamonds, $n = 2.7 \times 10^{11}$ cm$^{-2}$; triangles, $n = 2.1 \times 10^{11}$ cm$^{-2}$;
circles, $n = 1.6 \times 10^{11}$ cm$^{-2}$). {\bf (B)} Energy levels close to the $\nu = 3$ coincidence, showing
the field at which electrons transfer to the spin-up level. At the transition, the energy difference between
levels equals $\Delta E_{3,2}$. For clarity, only independent-electron energy levels are drawn. {\bf (C)}
Exchange-correlation energies $\Delta E_{\nu,l}$ extracted from (A) (with same symbols). {\bf (D)} Scaling of
$\Delta E_{\nu,l}$ with $l$, the number of occupied polarized energy levels.}
\end{figure*}

We now show that the positions of the peaks in ($B_\bot$, $B_{tot}$) phase space also lend themselves to a simple
interpretation in terms of a phase boundary between ferromagnetic states. We note that, as $\theta$ increases,
peaks appear alternatively within odd and even QH states. Following a standard argument used in {\it (5)}, we can
thus put the peaks in a one-to-one correspondence with the level-crossings of Fig.\ 1A. We label these crossings
with an index ($\nu$,$l$), where $\nu$ is the number of filled levels at the nearest $R_{xx}$ minimum and $l$ is
the rank of the coincidence, which we define as the number of fully polarized levels below the highest occupied
level (which undergoes the transition). In Fig.\ 3A, we plot the positions of the peaks for different densities
and tilt angles {\it (17)}.

Let us focus first on the (3,2) transition, near $\nu = 3$. In an independent-electron picture, where the only
energy contributions are from cyclotron and Zeeman energies (``bare'' energies), the transition between
unpolarized and polarized states happens exactly when the bare energy levels cross. By contrast, as shown in Fig.\
3B, interacting electrons may opt to transfer to the unoccupied level before the bare-level coincidence, i.e.\, at
a smaller $\theta$, in order to align their spins with the spins of the electrons in the lower levels [see {\it
(7)} for a similar argument]. Calling $\Delta E_{3,2}$ the energy difference between bare levels at the (3,2)
transition, we can write this difference as {\it (18)}
\begin{equation}
({5 \over 2}\hbar \omega_c - {1 \over 2}|g^*|\mu_B B_{tot}) - ({1 \over 2}\hbar \omega_c + {1 \over 2}|g^*|\mu_B
B_{tot})= \Delta E_{3,2}
\end{equation}
where $\omega_c=e B_\bot/m^*$. If we now write the total energy $E_{\nu,\pm 1/2}$ of electrons of spin $\pm 1/2$
in Landau level $\nu$ as a sum of their bare energies and of an exchange-correlation energy term $X_{\nu,\pm
1/2}$, we can then interpret $\Delta E_{3,2}$ as the difference $(X_{0,-1/2}-X_{2,+1/2})$, because the transition
takes place when the two competing energies $E_{0,-1/2}$ and $E_{2,+1/2}$ are equal. Using both Eq.\ 1 and our
data in Fig.\ 3A, we deduce the dependence of $\Delta E_{3,2}$ on $B_\bot$ at the transitions, which we plot in
Fig.\ 3C, using $g^* = 1.9$ and $m^* = 0.41m_e$. Although no accurate calculations for the exchange-correlation
energy terms in $\Delta E_{3,2}$ exist at present, we expect $\Delta E_{3,2}$ to be a fraction of the Coulomb
energy, $e^2/4\pi \epsilon l_B \simeq 65 B_{\bot}^{1/2}$, where $\epsilon \simeq 10 \epsilon_0$ is the dielectric
constant of AlAs and $l_B = (\hbar/eB_{\bot})^{1/2}$ is the magnetic length (the unit for $B_\bot$ is T and for
energy is K) {\it (19)}. In qualitative agreement with this expectation, the magnitude of the measured $\Delta
E_{3,2}$ increases with $B_\bot$. Our simple model described by Eq.\ 1 also explains the shift of resistance peaks
from high-$\nu$ to low-$\nu$ with increasing $\theta$. However, the dependence of $\Delta E_{3,2}$ on $B_\bot$ is
only slightly sublinear rather than proportional to $B_\bot^{1/2}$. A quantitative understanding of the magnitude
of $\Delta E_{3,2}$ and of its dependence on $B_\bot$ therefore requires future theoretical work, which includes
factors such as the finite layer-thickness of the 2D electrons, the mixing between Landau levels, and disorder.

We then define $\Delta E_{\nu,l}$, in analogy with $\Delta E_{3,2}$, as the difference between bare level energies
at the ($\nu$,$l$) transition, and plot $\Delta E_{\nu,l}$ in Fig.\ 3C. We note that all $\Delta E_{\nu,l}$
increase with $B_\bot$ and scale approximately as $l$, the number of occupied polarized levels. This scaling,
shown in Fig.\ 3D, is not surprising, because the $l$ polarized levels are the main levels contributing to $\Delta
E_{\nu,l}$, although not necessarily with equal weight.

The transition-induced magnetoresistance spikes we have described are apparent in all other samples we have
characterized so far and which contain similar carrier densities; these include one sample grown on a GaAs (311)B
substrate and five samples grown on (100), all of them 150\AA -wide AlAs quantum wells. On the other hand, we do
not observe the resistance spikes in high-density samples with only one subband occupied: In the density range
$3.5 \times 10^{11}$ to $5.2 \times 10^{11}$ cm$^{-2}$, where we have carefully monitored $R_{xx}$ as a function
of tilt angle, we observe that $R_{xx}$ minima do become slightly weaker near coincidence but do not exhibit any
additional peaks. At this point, we do not know why peaks appear only at low carrier densities.

We comment briefly on the resistance spikes themselves. We suggest that when the two competing ferromagnetic
states acquire the same energy, they separate into domains with opposite magnetizations (within the top Landau
level). Extended electron states then dissipate energy by scattering at the ferromagnetic domain walls, which
explains why the resistance rises at the transition. Indeed, theoretical studies of the QH liquid at $\nu = 1$
{\it (20)} hint that magnetic domains induced by a fluctuating Zeeman energy can cause dissipative transport. A
similar explanation has also been attributed to negative magnetoresistance in various ferromagnetic materials {\it
(21, 22)}. Nucleation sites for the domains may then be provided by topological quasiparticles akin to skyrmions,
which appear when two Landau levels cross {\it (23, 24)}. This may explain another conspicuous feature of our
hysteresis data in Fig.\ 2C, namely that the magnitude of the transition resistance is systematically higher in
the upward sweep than in the downward sweep, independent of the position of the peak with respect to the center of
the $R_{xx}$ minimum. This asymmetric trend might be due to the presence of a larger number of skyrmions as the
transition is approached from the low-$B$ side, i.e.\, from the state with the smaller magnetization.

In conclusion, we have demonstrated that a first-order Ising transition, when observed within the quantum Hall
regime, has a marked impact on the transport properties of a 2D electron gas. The resistance spikes we observe at
the transition can also shed light on the physics of magnetic domains in 2D electron systems. We have also learned
that AlAs quantum wells, which have been scarcely used in the past for measuring magnetotransport, can display
phenomena unobserved or unresolved so far in other semiconductor heterostructures. We thus hope that the many
structures we have yet to imagine on the basis of this compound, will provide a fertile ground for the observation
of new correlated electron phenomena.


\begin{references}
\bibitem{cooldowns}Data for Figs.\ 1B, 2, and 3 were collected during different cooldowns.
\end{references}
\end{document}